\documentclass[prb,reprint,showpacs,floatfix,superscriptaddress]{revtex4-1}
\usepackage[pdftex]{graphicx} 
\usepackage{dcolumn} 
\usepackage{bm} 
\usepackage{float}
\usepackage{amssymb} 
\usepackage{placeins}
\usepackage{amsmath} 
\usepackage{ulem} 
\usepackage{footmisc}
\usepackage[latin1]{inputenc}
\usepackage{tikz}
\usetikzlibrary{shapes,arrows}
 
\usepackage[hidelinks=true]{hyperref}

\begin{document}
\title{Advantageous nearsightedness of many-body perturbation theory contrasted with Kohn-Sham density functional theory}
\date{\today}
\author{J.\ Wetherell}
\affiliation{Department of Physics, University of York, Heslington, York YO10 5DD, United Kingdom}
\affiliation{European Theoretical Spectroscopy Facility}
\author{M.\ J.\ P.\ Hodgson}
\affiliation{Max-Planck-Institut f\"ur Mikrostrukturphysik, Weinberg 2, D-06120 Halle, Germany}
\affiliation{European Theoretical Spectroscopy Facility}
\author{L.\ Talirz}
\affiliation{Department of Physics, University of York, Heslington, York YO10 5DD, United Kingdom}
\affiliation{Institute of Chemical Sciences and Engineering, \'{E}cole Polytechnique F\'{e}d\'{e}rale de Lausanne, 1951 Sion, Switzerland}
\affiliation{European Theoretical Spectroscopy Facility}
\author{R.\ W.\ Godby}
\affiliation{Department of Physics, University of York, Heslington, York YO10 5DD, United Kingdom}
\affiliation{European Theoretical Spectroscopy Facility}

\begin{abstract} 
For properties of interacting electron systems, Kohn-Sham (KS) theory is often favored over many-body perturbation theory (MBPT) owing to its low computational cost. However, the exact KS potential can be challenging to approximate, for example in the presence of localized subsystems where the exact potential is known to exhibit pathological features such as spatial steps. By modeling two electrons, each localized in a distinct potential well, we illustrate that the step feature has no counterpart in MBPTs (including Hartree-Fock and $GW$) or hybrid methods involving Fock exchange because the spatial non-locality of the self-energy renders such pathological behavior unnecessary. We present a quantitative illustration of the orbital-dependent nature of the non-local potential, and a numerical demonstration of Kohn's concept of the nearsightedness for self-energies, when two distant subsystems are combined, in contrast to the KS potential. These properties emphasize the value of self-energy-based approximations in developing future approaches within KS-like theories.
\end{abstract}
\maketitle

Multiple approaches to the many-electron problem in quantum systems are available, each with strengths and weaknesses. Many-body perturbation theory (MBPT) is widely used for computing the electronic structure and properties of materials and molecules\cite{1367-2630-14-1-013056,PhysRev.139.A796,PhysRevB.75.235102,PhysRevLett.99.246403,PhysRevLett.74.1827,doi:10.1002/wcms.1344,RevModPhys.75.473,0953-8984-30-15-153002,doi:10.1063/1.4964690}, yet (approximate) Kohn-Sham density functional theory (KS-DFT) is often favored owing to its accuracy at a low computational cost\cite{HK64,KS65,DG,Burke12,RevModPhys.87.897}. The price for the computational efficiency is the difficulty in developing advanced approximations to the spatially local exchange-correlation (xc) potential of KS theory, $V_\mathrm{xc}(\textbf{r})$\cite{cohen2011challenges}. It has been noted that some modern approximate density functionals tend to focus on calculating accurate energies from empirical data to the detriment of the density\cite{Medvedev49}. In finite systems, reproducing the exact many-electron density requires $V_\mathrm{xc}(\textbf{r})$ to contain pathological features\cite{PhysRevLett.49.1691,PhysRevB.93.155146,vanLeeuwen1995}, which common approximations fail to capture\cite{PhysRevB.90.241107}. Thus practical calculations can be less reliable, e.g. for systems with strong localization such as molecules\cite{cohen2008insights}. In MBPT, on the other hand, exchange and correlation are described using a spatially non-local and energy-dependent potential, the self-energy operator. Generalized Kohn-Sham approaches\cite{PhysRevB.53.3764} have much in common with MBPT and are known to avoid some of the pathological aspects of KS theory insofar as quasiparticle energies are concerned\cite{PhysRevB.89.195134,Perdew2801}.

\begin{figure}[htbp] 
\centering
\includegraphics[scale=0.55]{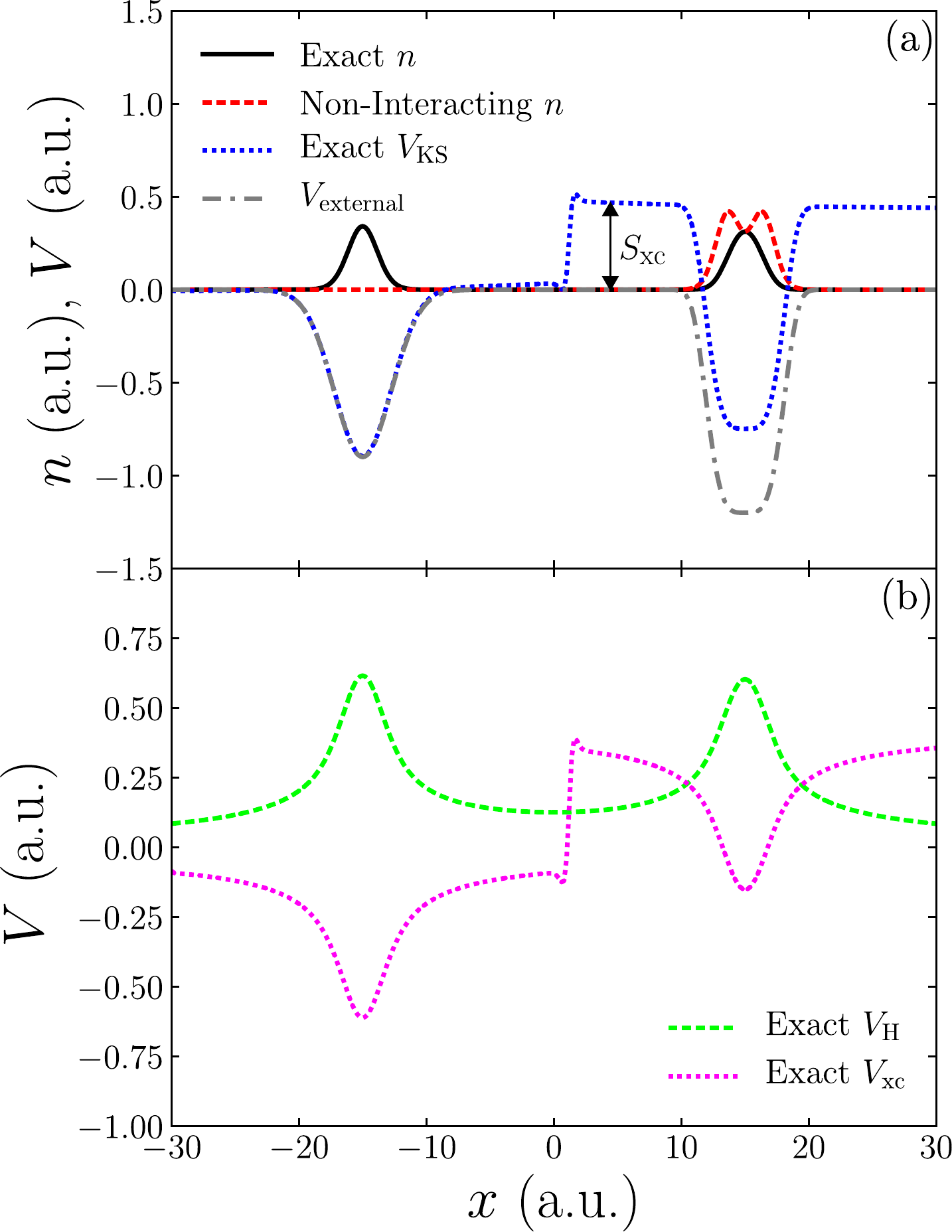} 
\caption{Exact many-electron density for two electrons in a 1D asymmetric double well and the corresponding exact KS system. (a) The exact many-electron and non-interacting density, the external potential and the exact KS potential. The external potential consists of two wells. The exact many-electron density corresponds to one electron in each well. Both of the non-interacting electrons occupy the right well. The exact KS potential for this system contains a step feature which raises the energy of the right subsystem by a constant, ensuring the correct distribution of electrons between wells. (b) The Hartree potential consists of two repulsive bumps centred one on each well, and the KS-xc potential is essentially the negative of the Hartree potential plus the step.}
\label{local}
\end{figure}

We describe the many-electron density in both \textit{exact} KS-DFT and two examples of MBPT for two interacting\footnote{We use a softened Coulomb repulsion $(|x-x'|+1)^{-1}$ as is appropriate in one dimension.} electrons in a 1D asymmetric double-well external potential (Fig.~\ref{local}(a)) for which a spatial step is known to be present in the exact KS potential\cite{PhysRevB.93.155146}. We use like-spin electrons in order to more closely capture the nature of exchange and correlation in larger systems, including the occupation of multiple spatial orbitals. We calculate the exact KS potential for this system by first solving the many-electron Schr\"odinger equation using our \texttt{iDEA} code\cite{PhysRevB.88.241102,PhysRevB.97.121102} in order to find the exact ground-state many-electron density. Then we `reverse-engineer' the KS equations to find the corresponding exact KS potential for this system, $V_\mathrm{KS}(x)$. 

Figure~\ref{local}(a) shows the exact many-electron density for our double-well system. The Coulomb repulsion between the electrons forces each electron in the system to localize in a distinct potential well. (In the absence of the Coulomb repulsion, both electrons would occupy the right well of the external potential as the lowest two non-interacting single-particle states of this system are localized in this well.) Figure~\ref{local}(a) shows the KS potential which yields the exact density for this system: a spatial step is present in the potential\footnote{While the KS potential shown yields the exact density to within computational precision, the localized nature of the two subsystems places a numerical limitation on the exact height of the step. However, analysis\cite{PhysRevB.93.155146} shows that it must be at least 0.03 a.u.}. The step acts to raise the right well by a constant relative to the left well. In doing so, the lowest energy state of the left well is made lower than the first excited energy state of the right well, and thus one occupied KS orbital is localized in the left well and the other is localized in the right well. This step feature has a non-local dependence on the density and is therefore beyond the capability of any common approximations\cite{PhysRevB.93.155146,PhysRevB.90.241107} to the KS-xc potential, such as the LDA\cite{KS65} or GGAs\cite{PhysRevLett.77.3865}.

The step in the KS potential is sharp owing to the large spatial separation of the potential wells in our system, which in turn means that the electrons are strongly localized. The step forms at the point in the density where the local effective ionization potential (IP) changes\cite{PhysRevB.93.155146}. This change occurs at the interface between the two individual potential wells (subsystems).

We now turn to the Hartree-Fock (HF) description of this system, the lowest level of MBPT, which already provides an excellent description of the electron density, even in the dissociation limit, as demonstrated in the Supplemental Material\footnote{For our like-spin electrons, Hartree-Fock correctly describes the localization of one electron in each well as the dissociation limit of the diatomic molecule is approached, while exhibiting the proper nearsightedness. For two electrons in a singlet state, however, this is not the case for normal (restricted) HF, as it neglects static correlation \cite{Cohen792}, and it would be necessary to use unrestricted HF \cite{PhysRev.102.1303} (in which electrons with different spins occupy different orbitals), or to a higher level of many-body perturbation theory, in order to achieve a similar combination of accuracy and nearsightedness \cite{doi:10.1021/ct200345a}.}.

\begin{figure}[htbp]
\centering
\includegraphics[scale=0.55]{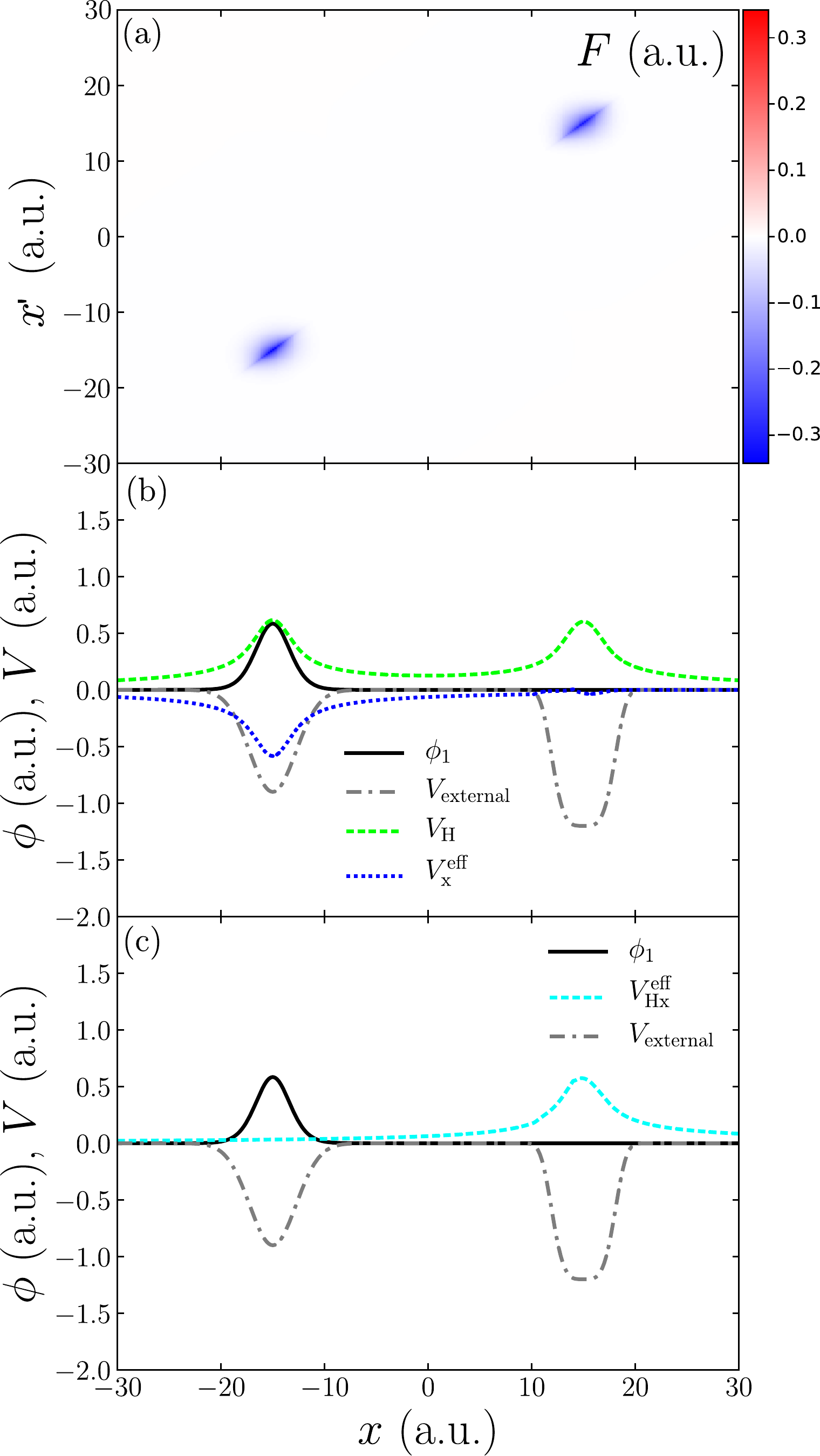} 
\caption{The double well described by HF theory. (a) The Fock operator, which yields a highly accurate density (see Supplemental Material). The pathological features in the exact KS potential are absent. (b) Effective potentials (see text) experienced by $\phi_1$. It `feels' the external potential $V_{\mathrm{ext}}(x)$, the Hartree potential $V_{\mathrm{H}}(x)$ of the whole system, which consists of a repulsion bump from both orbitals ($\phi_0$ and $\phi_1$) and its own effective exchange potential $V^{\mathrm{eff}}_{\mathrm{x},1}(x)$ that acts to cancel the Hartree potential due to its own presence (SIC). (c) The overall effective Hartree-xc potential felt by orbital $\phi_1$ -- the electron in the left well feels the repulsion of the electron in the right well and vice-versa resulting in a density corresponding to one electron in each well as per the many-electron density.}
\label{non_local}
\end{figure}

In Fig.~\ref{non_local}(a) we show the Fock operator for this system: no features corresponding to the step in the KS potential are visible and the operator appears to have an approximately local dependence on the density. $F$ is seen to be non-local on the length scale of the subsystem, but \textit{not} of the overall system. Our analysis of the nearsightedness of the Fock operator below constitutes a quantitative confirmation of these observations.

The concept of an effective orbital-dependent local potential is illustrative. For a particular orbital, the mathematical effect of a non-local potential is exactly equivalent of an effective local potential; in the case of the exchange operator this is
\begin{equation}
\label{eq:eff}
V^{\mathrm{eff}}_{\mathrm{x},m}(x) = \frac{1}{\phi_m(x)}\int F(x,x')\phi_m(x')dx'.
\end{equation}
It is key to note that this effective potential is different for every orbital, in contrast to KS theory in which every electron feels the same local effective potential.

Figure~\ref{non_local}(b) shows the effective potentials felt by $\phi_1(x)$ in HF theory\footnote{To handle the singularity arising from the node in $\phi_1(x)$ which occurs in the vicinity of the right well, a careful numerical treatment of the denominator of Eq.~\ref{eq:eff} is necessary.}. This orbital is localized in the left well. It feels the external potential and the Hartree potential of the \textit{whole} system, which consists of two large positive bumps; one is in the region of the left well and the other the right well. In addition, $\phi_1$ feels its effective local exchange potential, which acts to cancel out the Hartree potential on the left, i.e., the self-interaction correction (SIC), but is negligible on the right; see Fig.~\ref{non_local}(b). Figure~\ref{non_local}(c) shows the resulting net potential felt by $\phi_1$: the left electron feels the Coulomb repulsion due to the right electron, ensuring that each electron occupies its own well in accordance with the many-electron picture. Thus, for weakly correlated system, such as this one, HF successfully localizes electrons. For systems comprising more complex separated subsystems, further vertex corrections beyond $GW$ can be significant, but nearsightedness should remain assured by the self-energy diagrams' analytic dependence on the single-particle orbitals, with terms connecting the two systems going to zero at large separations.

It is straightforward for the Fock operator to remove the self-interaction (SI) part of the Hartree potential \textit{for each electron separately} owing to to its spatial non-locality. In contrast, the spatially local exact xc potential does not have such freedom, and must remove the SI part of the Hartree potential for all electrons simultaneously, this acts to essentially cancel the \textit{whole} Hartree potential; see Fig.~\ref{local}(b). Thus, without each electron experiencing the Hartree potential due to the other electron, the KS potential must instead include a spatial step at the interface between the electrons.

To demonstrate that the Fock operator of the \textit{whole} double-well system consists of the SIC for each electron and \textit{no additional features} we calculate the Fock operator for each one-electron subsystem completely independently ($F_L$ and $F_R$). For the
case shown in Fig.~\ref{non_local}, $F_L + F_R$ reproduces the Fock operator for the composite system to high accuracy\footnote{See Supplementary Material.}:
$\sim0.03$ a.u.\ ($\sim 2\%$ of the scale on which $F$ varies)\footnote{The density calculated from employing $F_L + F_R$ in the HF equations for the whole system yields a density which is very similar to the true HF density for this system:
$\sim0.01$ a.u. ($\sim 2\%$ of the scale on which $n$ varies). This value approaches zero as the wells are separated.}, i.e.,
\begin{equation} \label{eq:Fsum}
F(x,x') = F_L(x,x') + F_R(x,x').
\end{equation}
Equation~\ref{eq:Fsum} becomes exact in the limit that the subsystems are infinitely separated. 

This property of $F$ (and more generally the self-energy in MBPT) is an example of Kohn's `nearsightedness' principle\cite{PhysRevLett.76.3168}, in which the physical properties of one subsystem are blind to those of another, distant, subsystem. In contrast to the self energy, the exact KS potential does \textit{not} exhibit this `nearsightedness' principle. The exact $V_{\mathrm{xc}}(x)$ for the left subsystem is simply the negative of the Hartree potential, and the same for the right subsystem. Therefore their sum does not reproduce the KS potential for the whole system as this contains the step at the interface between the subsystems, i.e.,
\begin{equation}
V_{\mathrm{xc}}(x) =V^L_{\mathrm{xc}}(x) + V^R_{\mathrm{xc}}(x) + S_{\mathrm{xc}}(x).
\end{equation}
This highlights the straightforward nature of a non-local potential compared to a local potential.

\begin{figure}[htbp] 
\centering
\includegraphics[scale=0.55]{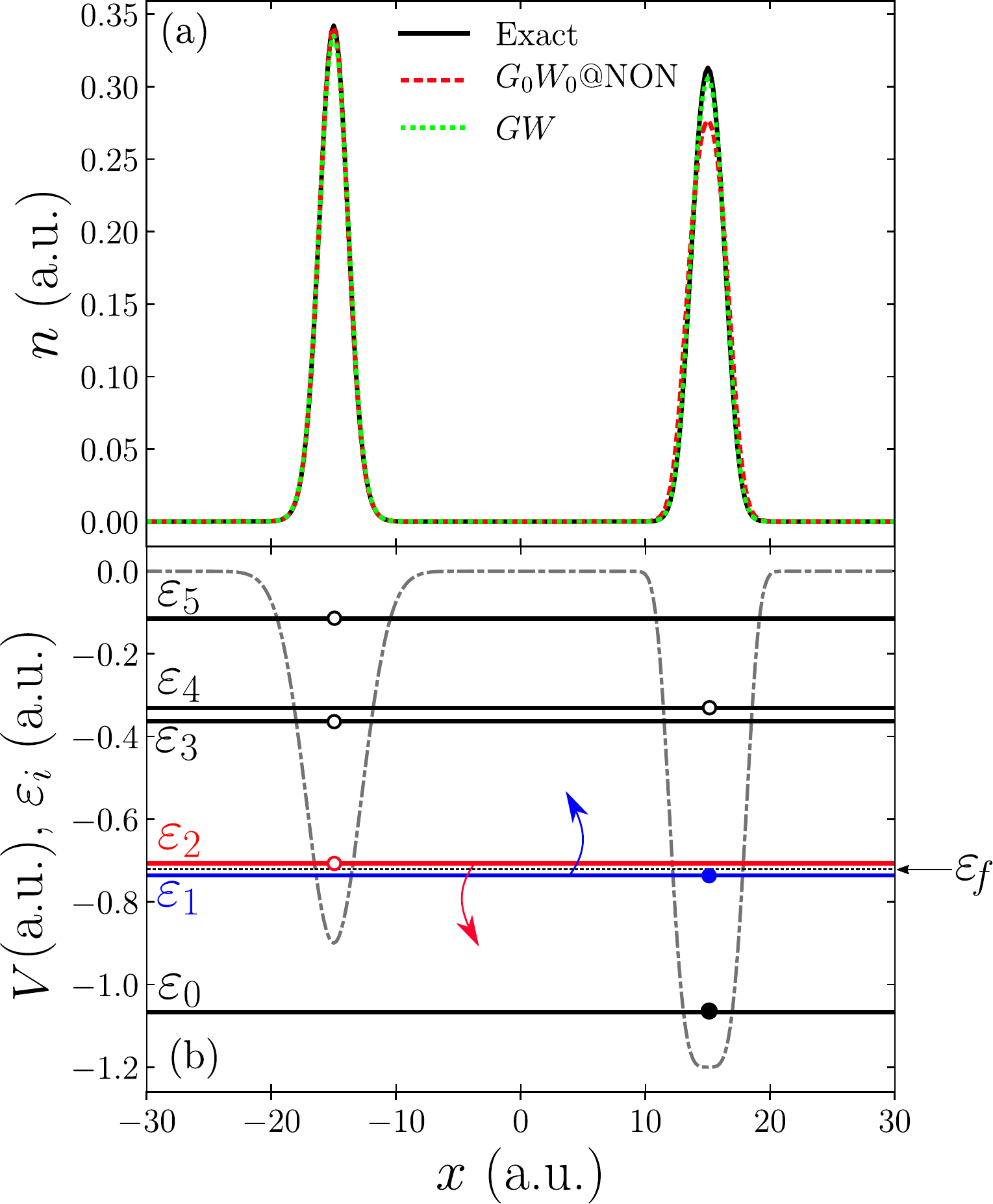} 
\caption{$GW$ calculations for the double-well system. (a) One-shot $G_0W_0$ density starting with non-interacting (NON) orbitals yields a surprisingly accurate density, although the shape of the density in each well is broadened relative to the exact. The fully self-consistent $GW$ density is much more accurate, as this broadening is significantly reduced. (b) Orbitals are swapped in the one-shot case. The horizontal lines ($\varepsilon_0$ and $\varepsilon_1$) indicate the non-interacting single-particle starting energies, where the circle on the line indicates in which of the two wells a particular orbital occupies -- a filled circle indicates an occupied state and a hollow circle represents an unoccupied state. Initially both of the occupied orbitals are localized in the right well, thus giving the non-interacting density; see Fig.~\ref{local}(a). In the first iteration of both $GW$ calculations the orbitals of the self-energy are swapped; the HOMO orbital (blue) is raised above the Fermi energy, and the LUMO (red) is brought below. This means that after the swap one electron occupies the left well and the other the right, as required.}
\label{gw}
\end{figure}

We now move onto the $GW$ approximation, the next level of MBPT. First, we demonstrate that the density from a one-shot ($G_0W_0$) calculation is surprisingly accurate even when starting from a set of orbitals which yield a very poor initial density. In our case we choose to start from the non-interacting orbitals of the external potential which yield a density that is quantitatively different from the many-electron density; see Fig.~\ref{local}(a). As shown in Fig.~\ref{gw}(a), the $G_0W_0$ correctly gives one electron in each well in contrast to its starting point, but the shape of the density in the subsystems is broadened relative to the exact. We find that the self-energy swaps occupied and unoccupied starting orbitals\footnote{We use first order perturbation theory when computing the updated quasi-particle energies from the self-energy.} when the Dyson equation is solved as shown in Fig.~\ref{gw}(b). As the $G_0W_0$ case has a non-local potential these orbitals can be moved independently and thus the self-energy needs no step feature. This is in contrast to the KS case where all of the orbitals in the right well are shifted simultaneously by the magnitude of the step in order to get the correct occupation of KS electrons. This swapping mechanism implies that the accuracy of $G_0W_0$ depends on the features of the unoccupied as well as the occupied starting orbitals.

Second, we perform a fully self-consistent $GW$ calculation for this system. The fully self-consistent $GW$ density is very accurate, albeit slightly worse than HF. This small error in this density is due to the self-screening error\cite{PhysRevA.75.032505,PhysRevB.85.035106,2009PhRvA..79e2513F,1402-4896-86-6-065301,PhysRevB.85.035106,1402-4896-86-6-065301}, which arises from a spurious non-zero correlation part of the self-energy (see Supplemental Material). We recently demonstrated\cite{PhysRevB.97.121102} that the self-screening error may be accurately corrected by a local-density-type expression which therefore retains $\Sigma_{\mathrm{xc}}$'s nearsighted character within each well.

\begin{figure}[htbp] 
\centering
\includegraphics[scale=0.55]{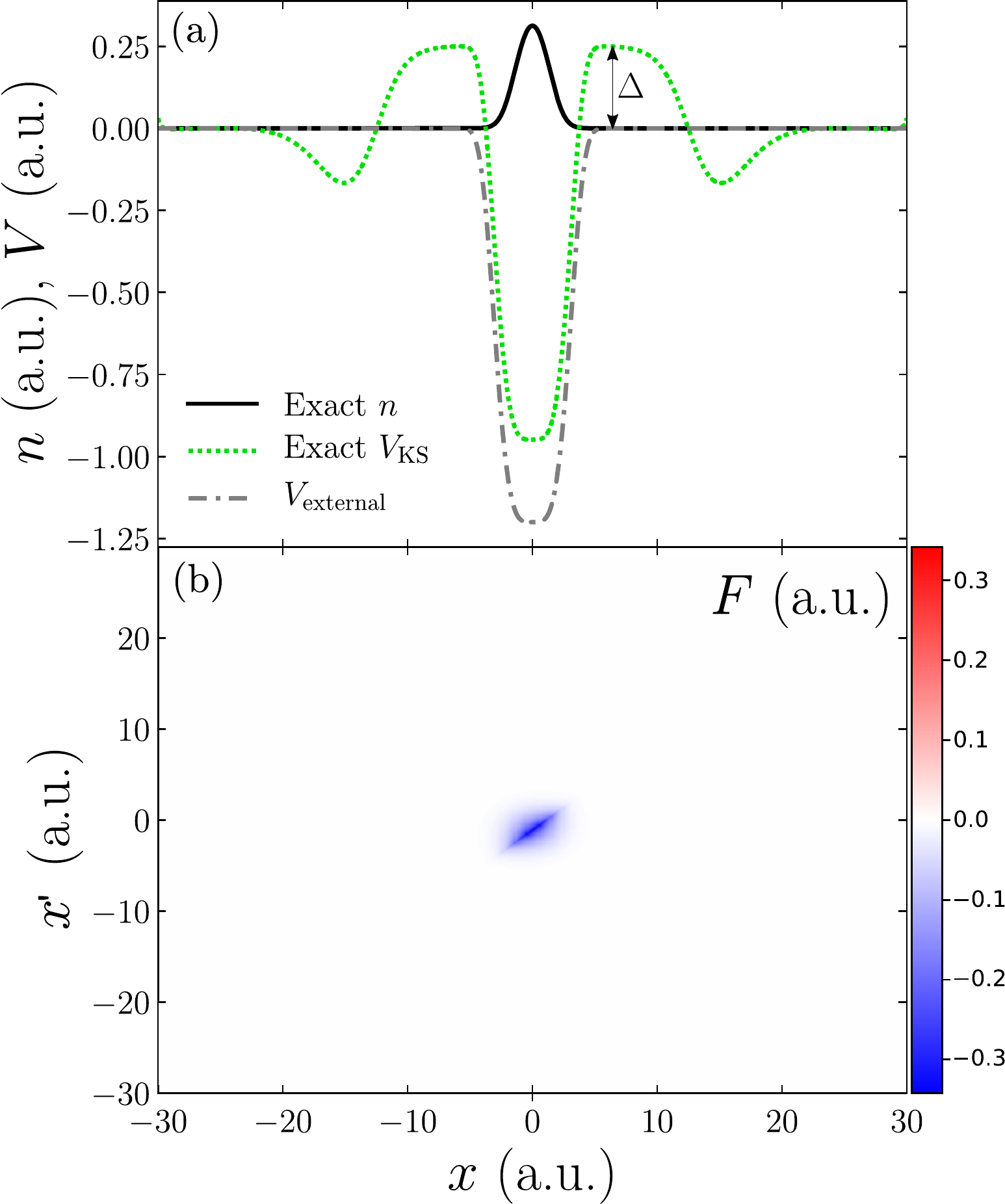} 
\caption{(a) The right-hand well of our double-well system with $1.0001$ electrons. The exact electron density is indistinguishable from the HF density. The exact KS potential has a plateau in the vicinity of the well with height $\Delta$. This plateau occurs in the KS system as a result of the derivative discontinuity. (b) The corresponding Fock operator contains no features which correspond to the steps in the KS potential.}
\label{dd}
\end{figure}

We also consider an open system, connected to an electron reservoir, allowing a fractional number of electrons. The exact KS potential experiences a jump by a spatially constant shift $\Delta$ when the number of electrons, $N$, in the system infinitesimally surpasses an integer\cite{PhysRevLett.49.1691}. This is known as the `derivative discontinuity' as it is a result of the discontinuity in the derivative of the total energy as a function of $N$. It is essential in KS theory if one wishes to determine the electron affinity (EA) from the single-particle KS energies of a system alone, yet it is not reproduced at all by common approximations\cite{perdew1984can,cohen2011challenges}.

We now model only the right-hand well of our double-well system; see Fig.~\ref{dd}(a). We investigate what happens to the non-local potential of HF when $\delta=10^{-4}$ of an electron is added to a one-electron system. First we calculate the exact density for the $1+\delta$-electron system and the corresponding exact KS potential; see Fig.~\ref{dd}(a). When $\delta$ is small but finite, the shift $\Delta$ is a no longer uniform throughout all space but a plateau -- it is uniform in the center but at each side has a step in the potential; see Fig.~\ref{dd}(a). The height of these steps is the discontinuity $\Delta$. In the limit that $\delta \rightarrow 0^+$, these steps form further and further away from the well and hence the plateau becomes a spatially uniform shift\cite{doi:10.1021/acs.jpclett.7b02615}.

The Fock operator corresponding to the $1+\delta$-electron system is shown in Fig.~\ref{dd}(b). The steps in the KS potential do not correspond to any features in this Fock operator, and thus do not occur in the effective exchange potentials either. Instead, when $\delta$ of an electron is added to the HF system it experiences a different effective potential to the one felt by the whole electron which already occupies the well. The additional fraction of an electron experiences essentially just the Hartree potential of the whole electron originally in the system plus the external potential; whereas the whole electron in the system feels effectively no Hartree potential from $\delta$. Thus $\delta$ has a higher energy than the other electron in the system which in turn determines the system's new IP without the need for any discontinuous change to the Fock operator. This reasoning also applies to the case for hybrid density functionals (which combine the Fock operator with a usually local xc potential)\cite{doi:10.1063/1.472933,PhysRevMaterials.2.040801,doi:10.1063/1.2187006} and other schemes within generalized KS theory\cite{PhysRevB.53.3764,Perdew2801} as well as the $GW$ approximation, all of which are known to yield improved values for the fundamental gap (IP minus EA) compared to (approximate) Kohn-Sham gaps\cite{doi:10.1063/1.472933,PhysRevMaterials.2.040801}.

In conclusion, a quantitative analysis of orbital-dependent effective potentials and the nearsightedness of the self-energy operator shows that the crucial pathological features of the exact Kohn-Sham exchange-correlation potential --  beyond the capability of common approximations -- are not required in the non-local potential of many-body perturbation theory: in effect, each electron is able to experience a different local potential. This emphasizes the potential value of constructs from self-energy methods in developing future approaches within KS-like theories.

\begin{acknowledgments}
We acknowledge funding from the York Centre for Quantum Technologies (YCQT), and thank Eli Kraisler for helpful comments.
\end{acknowledgments}

\bibliography{main.bib}
\end{document}